\documentclass[a4paper]{jpconf}

\def\AEF{Faraggi A E}

\usepackage{amsmath,amssymb,lmodern}
\usepackage{graphicx}
\usepackage{amsfonts}
\usepackage{bm}

\def\ibid#1#2#3{{\it ibid} {\bf #1}, (#2) #3}

\def\NPB#1#2#3{#2 {\it Nucl.\ Phys.}\/ {\bf B#1} #3}
\def\PLB#1#2#3{#2 {\it Phys.\ Lett.}\/ {\bf B#1} #3}

\def\PRD#1#2#3{#2 {\it Phys.\ Rev.}\/ {\bf D#1}  #3}
\def\PRC#1#2#3{{\it Phys.\ Rev.}\/ {\bf C#1} (#2) #3}
\def\PRL#1#2#3{#2 {\it Phys.\ Rev.\ Lett.}\/ {\bf #1} #3}

\def\MODA#1#2#3{#2 {\it Mod.\ Phys.\ Lett.}\/ {\bf A#1} #3}

\def\IJMP#1#2#3{#2 {\it Int.\ J.\ Mod.\ Phys.}\/ {\bf A#1} #3}

\def\EJP#1#2#3{#2 {\it Eur.\ Phys.\ Jour.}\/ {\bf C#1} #3}
\def\JHEP#1#2#3{#2 {\it JHEP}\/ {\bf #1} #3}

\def\etal{{\it et al\/}}

\newcommand{\beq}{\begin{equation}}
\newcommand{\eeq}{\end{equation}}
\newcommand{\beqa}{\begin{eqnarray}}
\newcommand{\beqn}{\begin{eqnarray}}
\newcommand{\eeqn}{\end{eqnarray}}
\newcommand{\eeqa}{\end{eqnarray}}

\usepackage{epsf}
\usepackage{graphicx}
\begin{document}
\title{Spinor--vector duality and sterile neutrinos \\
in string derived models}

\author{Alon E. Faraggi}

\address{ Department of Mathematical Sciences, University of Liverpool, 
Liverpool L69 7ZL, UK}

\ead{alon.faraggi@liv.ac.uk}

\begin{abstract}

The MiniBooNE collaboration found evidence for the existence 
of sterile neutrinoS, at a mass scale comparable to the active 
left--handed neutrinos. While sterile 
neutrinos arise naturally in large volume string scenarios, they 
are more difficult to accommodate in heterotic--string derived models 
that reproduce the GUT embedding of the Standard Model particles.
Sterile neutrinos in heterotic--string models imply the existence 
of an additional Abelian gauge symmetry at low scales, possibly 
within reach of contemporary colliders. I discuss the construction 
of string derived $Z^\prime$ models that utilise the spinor--vector 
duality to guarantee that the extra $U(1)_{Z^\prime}$ symmetry can remain
unbroken down to low scales. 

\end{abstract}

\section{Introduction}

Physics is first and foremost an experimental science. Be that as it may,
the language that is used to parametrise experimental data is mathematics. 
Formulating mathematical models that can encompass a broader range of 
physical data is therefore a well defined scientific problem.

All contemporary experimental observations are accounted for by two 
fundamental theories. At the celestial, galactical and cosmological 
scales Einstein's theory of general relativity reigns supreme. At 
the atomic and subatomic scales the Standard Model of particle physics
accounts for all experimental data to an impressive and increasing precision. 
Yet the two theories are mutually incompatible. It is therefore mandated 
to seek mathematical frameworks that can accommodate the two theories in 
a mathematically consistent framework.

From an experimental point of view what is required is the continued 
precision measurement of prevailing mathematical parameters of 
contemporary experiments. These include: more precise measurements of 
the scalar sector and its couplings; more precise measurements of the 
neutrino sector and of its particle content; more refined measurement
of the Cosmic Microwave Background Radiation; elucidation of the origin
of the highest energy cosmic rays; further refined measurements of the 
cosmic microwave background radiation; further discovery and measurements
of gravitational waves. Questions abound. Further afield can be listed the 
detection of the premordial cosmic neutrino and gravitational field 
backgrounds. 
Extending the energy threshold should be sought to test contemporary
theories in new regimes and to explore possible signals that go beyond
the current paradigms. Current theoretical extensions of the Standard 
Models do not provide an unambiguous indication for where experiments should 
look for detection of their signatures. In the absence of such 
unambiguous theoretical guidance experimentalists should focus 
on refined tests of the existing frameworks. The experimental 
successes of the past decades have enabled an unprecedented level of 
understanding of the basic constituents of matter and interactions.
Future world wide experimental facilities will extend the era of exploration,
that started with Galileo's telescope, to new horizons.

One exciting experimental result that requires resolution is the recent
report by the MiniBooNE collaboration for further evidence for the 
existence of sterile neutrinos \cite{miniboone}.
Evidence for sterile neutrinos was reported by the LSND
collaboration in 1993 \cite{lsnd}. While the experimental 
and phenomenological analysis is far from settled \cite{dentler}, if the 
result is strengthened by future experimental data, it will
have profound implications on string phenomenology. While
sterile neutrinos may be naturally accommodated in large 
volume string scenarios, they are much harder to reconcile 
with the more favoured string GUT models. 

From a theoretical perspective, contemporary interest is 
in the synthesis of quantum mechanics and general relativity.
Observational phenomena at the highest energy and shortest 
distance scales probed to date, point to the Standard Model of 
particle physics as the viable theory. The observation 
of a scalar particle at the LHC is compatible with
the Standard Model Higgs particle. 
All the observed data to date
is consistent with the hypothesis that the Standard Model remains 
a viable perturbative parametrisation up to very high energy scales, 
{\it e.g.} the Grand Unified Theory (GUT), where all the gauge 
interactions are of comparable strength, or the Planck scale, 
where the magnitude of the gauge and gravitational couplings is comparable.
Further evidence for this hypothesis stems from the logarithmic
running of the Standard Model parameters, and the suppression
of left--handed neutrino masses and proton decay mediating operators. 
Most tantalising is the
fact that the Standard Model gauge charges of each family fit into
spinorial {\bf16} representation of $SO(10)$.
The Standard Model gauge charges are experimental observable
that were determined in the process of the experimental 
discovery of the Standard Model. Since the Standard Model 
matter sector consist of three generations; three group factors; 
and six left--handed Weyl multiplets (including the right--handed
neutrino), fifty four independent parameters are required to account 
for the Standard Model gauge charges. The embedding of the Standard 
Model matter spectrum in  $SO(10)$ representations reduces this number
to one, {\it i.e.} the required 
number of spinorial {\bf 16} spinorial representations of $SO(10)$. 
An amazing coincidence indeed! However, the saga is not complete.
Gravity exists! Without it we will all be floating up in space.
String theory enables
exploration of the augmentation of the Standard Model with gravity
in a perturbatively consistent framework. Furthermore, while 
in the context of point quantum field theories the Standard Model
may be augmented with an infinite number of continuous parameters, 
string theory severely constrains the allowed extensions, 
and may shed light on the origin of its existing parameters.
A natural question is how can sterile neutrinos be accommodated 
in string vacua \cite{sninsv}? 

\section{Sterile neutrinos in large volume scenarios}

Sterile neutrinos may be naturally accommodated in 
large volume scenarios. As depicted in the 

\vspace{5mm}
\centerline{
{\epsfxsize=5cm\leavevmode\epsffile[0 0 471 364]{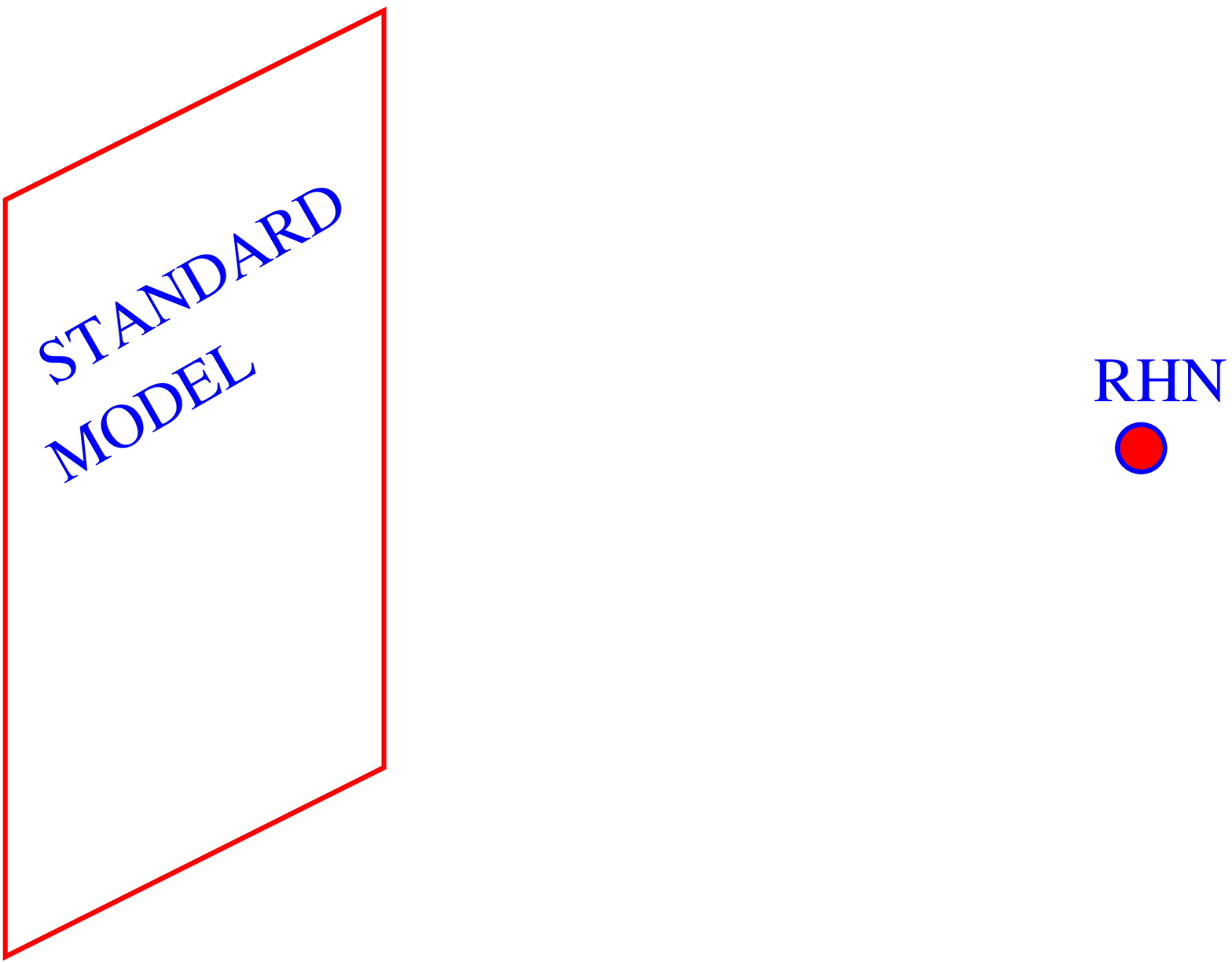}}}
\noindent
figure. 
In these scenarios the Standard Model states are confined to an observable 
brane, whereas the right--handed neutrino field is a bulk field.
The couplings of these fields to the Standard Model states 
are then suppressed by the volume of the extra dimensions, 
very much like the suppression of the gravitational couplings
\cite{lvneutrinos}. Assuming a five dimensional scenario,
$(x^\mu,y)$ where $\mu=0,\cdots,3$ and a $y$ is a compactification 
circle with radius $R$.
The Dirac spinor field in the bulk decomposes in the Weyl basis as
$\Psi = \left( \nu_R , \bar{\nu^c}_R  \right)$,
and has the usual Fourier mode expansion
\begin{equation}
\nu_R^{(c)}(x,y) = \sum_n \frac{1}{\sqrt{2 \pi r}} \nu_{Rn}^{(c)}(x) e^{iny/r}
\end{equation}
The mass spectrum in four dimensions then contains
a tower of Kaluza--Klein states with Dirac
masses $n/r$ and a free action for the lepton
doublet, localized on the wall. 
The leading interaction term between the bulk and wall fermion fields is
\begin{equation}
S^{\rm{int}} = \int d^4 x \lambda l(x) h^*(x) \nu_R(x,y=0)
\label{intf}
\end{equation}
with $\lambda$ being a dimensionless parameter.
The Yukawa coupling $\lambda$ to all brane fields is rescaled like the 
dilaton and graviton couplings. 
The effective Yukawa on the four dimensional brane is given by
\begin{equation}
\lambda_{(4)}=\frac{\lambda}{\sqrt{r^nM_*^n}},
\end{equation}
which leads to very strong suppression of the Dirac mass. 
Thus, in large volume scenarios the bulk fields behave 
like sterile neutrinos with suppressed couplings to the 
left--handed neutrino brane fields with potentially 
observational effects \cite{lvneutrinos}. In this scenarios, however, 
the appealing embedding of the Standard Model states in $SO(10)$
representations is abandoned. 

\section{String GUT models in the free fermionic construction}

String phenomenology is in some respect an answer in search of a question. 
The answers are provided in the form of the Standard Models of particle physics
and cosmology and their BSM extensions. One question that one may ask is what is
the true string vacuum that reproduces the detailed features of these models, with
possibly some additional predicted signatures beyond the Standard Models.
Another question that one may ask is whether one can identify generic 
signatures of classes of string compactifications in the experimental 
particle and cosmological data. In this talk I argue that the existence of 
light sterile neutrinos in heterotic string GUT vacua mandates the 
existence of a light $Z^\prime$ vector boson, which in turn is hard to 
incorporate in these models. 

The perturbative and non--perturbative duality symmetries among ten dimensional
string theories, as well as eleven dimensional supergravity \cite{ht}
reveals, as depicted in the figure below,

\vspace{5mm}
\centerline{
{\epsfxsize=5cm\leavevmode\epsffile[0 0 538 532]{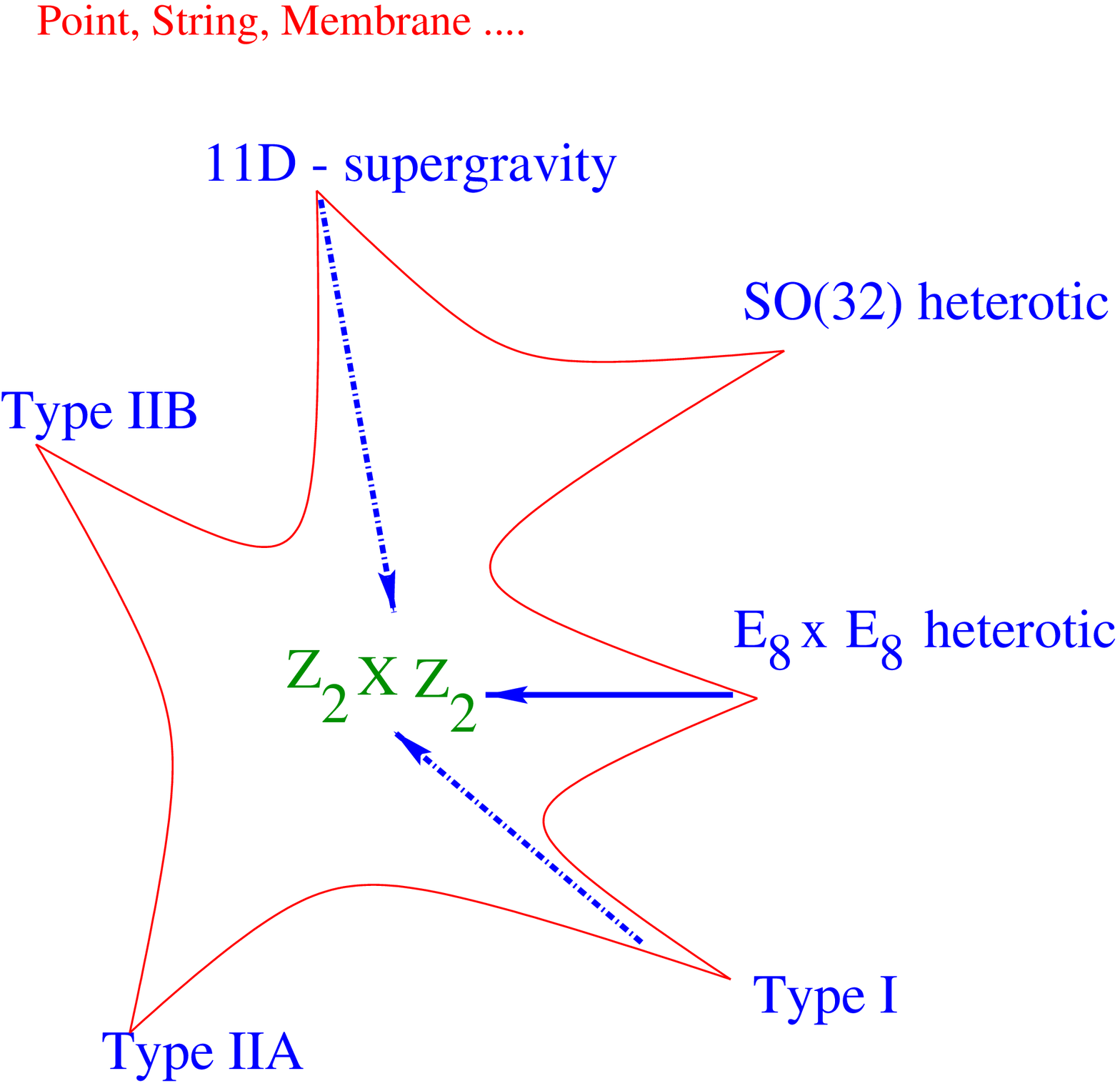}}}
\vspace{5mm}
\noindent
that the different string theories are limits of 
a more fundamental theory. 
If we take the embedding of the Standard Model states in $SO(10)$ 
representation as the primary guide in seeking phenomenological 
signatures of string theory, then the string limit that should be
used is the ten dimensional $E_8\times E_8$ heterotic--string theory,
because it is the only  
string theory that produces spinorial $SO(10)$ representations
in its perturbative spectrum. In this respect we should anticipate
that working in a perturbative limit of the more fundamental 
theory reveal some properties of the true string vacuum, but cannot
fully characterise it. It may well be that other features 
can only be gleaned in a different limit. An example at hand is the 
stablisation of the dilaton moduli that determines the strength of the 
gauge and gravitational interactions. In the perturbative heterotic 
limit the dilaton has a run away behavior and cannot be stablised at 
a finite value. The underlying picture depicted in the figure 
above reveals that the dilaton can be interpreted as a moduli 
of an eleventh dimensions. Thus, to stabilise the dilaton at a 
finite value necessitates moving away from the perturbative 
heterotic limit. This illustrates the point that we should not
expect any of the limits to fully describe the true string vacuum, 
but at best to capture some of its properties. On the other hand the
true string vacuum will have some characteristics that are independent 
of the limit around which we expand. What those features are is a matter
of research and debate, but we may hypothesise that they might be related
to the underlying structure of the internal compaticified space. This
is suggested in the figure that exemplifies this notion with the case 
study of the $Z_2\times Z_2$ toroidal orbifold in the different limits. 

Phenomenological studies of string GUT theory proceed by compactifying
the heterotic string from ten to four dimensions. The degrees of freedom 
of the internal six dimensional space can be represented as 
free or interacting world sheet conformal field theories. In simple cases 
the effective low energy quantum field theory is represented in the form
of supergravity compactified on a complex internal manifold. However, at 
present our basic understanding of string theories is rudimentary and
phenomenological studies should be regarded as exploratory. With this 
caveat in mind, simple classes of compactifications are orbifolds 
of six dimensional toroidal spaces.
Of those the simplest orbifold is the $Z_2$ orbifold of a one dimensional 
torus, {\it i.e.} a circle. Taking the internal space to be a six torus, 
consistency dictates that a $Z_2$ orbifold can act on four internal 
coordinates at a time. The initial toroidal compactification
gives rise to $N=4$ spacetime supersymmetry and each $Z_2$ orbifold 
reduces the number of supersymmetries by two. Hence, to reduce the 
number of spacetime supersymmetries requires a $Z_2\times Z_2$ orbifold
of the six dimensional internal toroidal space. This class of string 
compactifications is among the simplest that one may construct, and it
possesses several other appealing properties. One being that it naturally gives 
rise to three generations \cite{nahe}. 
Heuristically, we may attribute the emergence 
of three generations in the $Z_2\times Z_2$ orbifolds to the fact that we are
dividing an internal six dimensional space into factors of two. 
The $Z_2\times Z_2$ orbifold contains three twisted sectors and in 
many of the three generation models each one of the twisted sectors produces 
one generations. Another important property of the phenomenological 
$Z_2\times Z_2$ orbifold models is that they preserve the $SO(10)$ embedding
of the weak hypercharge, hence facilitating the construction of string GUT models.
The $Z_2\times Z_2$ orbifold produces an abundance of phenomenological 
three generation models that serves as a case study to explore possible 
relations between string theory and experimental data. 

The $Z_2\times Z_2$ orbifold models have been most amply studied by using the 
free fermionic formulation of the heterotic-string in four dimensions. The 
equivalence of two dimensional fermions and bosons entails that this 
formulation is entirely equivalent to the toroidal orbifold construction. 
Indeed, for any free fermion model one can find the bosonic equivalent 
\cite{panos}.
Since the late eighties the free fermion models served as a laboratory to 
study phenomenological issues in the Standard Model and unification, including:
the construction of string models that produced solely the spectrum 
of the Minimal Supersymmetric Standard Model in the observable 
charged matter sector \cite{fny}; fermion mass hierarchy and mixing 
\cite{fmhandm}; neutrino masses \cite{nmasses}; and more \cite{more}.
The $Z_2\times Z_2$ orbifold models represent a case study and other approaches
are being pursued \cite{others}. The task of string phenomenology is to develop
the tools to discern between the different cases and identify their signatures 
in the experimental data.

In the fermionic formulation of the heterotic--string in four dimensions
all the worldsheet degrees of freedom required to cancel the conformal anomaly 
are represented in terms of two dimensional free fermions
on the string worldsheet. In the standard notation the 64 worldsheet 
fermions the lightcone gauge are denoted as: 
\leftline{~~~${\underline{{\hbox{Left-Movers}}}}$:~~~~~~~~~~~~~~~~~~~~~~~~
~~~~$\psi^\mu,~~{ \chi_i},~~{ y_i,~~\omega_i}~~~~(\mu=1,2,~i=1,\cdots,6)$}
\vspace{4mm}
\leftline{~~~${\underline{{\hbox{Right-Movers}}}}$}
$${\bar\phi}_{A=1,\cdots,44}=
\begin{cases}
~~{ {\bar y}_i~,~ {\bar\omega}_i} & i=1,{\cdots},6\cr
  & \cr
~~{ {\bar\eta}_i} & i=1,2,3~~\cr
~~{ {\bar\psi}_{1,\cdots,5}} & \cr
~~{{\bar\phi}_{1,\cdots,8}}  & 
\end{cases}
$$
Of these the $\{y,\omega\vert{\bar y},{\bar\omega}\}^{1,\cdots,6}$
correspond to the 
six compactified dimensions of the internal manifold; 
${\bar\psi}^{1,\cdots,5}$ generate the $SO(10)$ GUT symmetry; 
${\bar\phi}^{1,\cdots,8}$ generate the 
hidden sector gauge group; and ${\bar\eta}^{1,2,3}$ 
generate three $U(1)$ symmetries.
Models in the free fermionic formulation are written in terms of boundary 
condition basis vectors, which specify the transformation properties 
of the fermions around the noncontractible loops of the worldsheet torus, 
and the Generalised GSO projection coefficients of the one loop 
partition function \cite{fff}. The free fermion models correspond to 
$Z_2\times Z_2$ orbifolds with discrete Wilson lines \cite{panos}.

\section{Sterile neutrinos in free fermion models} 

In the heterotic--string GUT models the $SO(10)$ gauge symmetry is broken 
at the string scale to one of its subgroups by discrete Wilson lines. Due
to the underlying $SO(10)$ symmetry the models possess some common features, 
irrespective of the unbroken $SO(10)$ subgroup, The neutrino mass
matrix in these models takes the generic form 
\begin{equation}
{\left(
\begin{matrix}
                 {\nu_i}, &{N_i}, &{\phi_i}
\end{matrix}
   \right)}
  {\left(
\begin{matrix}
         0   &       (M_{_D})_{ij}    &             0                 \\
  (M_{_D})_{ij}&          0            & \langle{\overline {\cal N}}\rangle_{ij} \\
          0  &\langle{\overline{\cal N}}\rangle_{ij} & \langle\phi\rangle_{ij} \\
\end{matrix}
   \right)}
  {\left(
\begin{matrix}
                 {\nu_j}  \cr
                 {N_j}\cr
                 {\phi_j} \cr
\end{matrix}
   \right)},
\label{nmm}
\end{equation}
where $M_{_D}$ is the Dirac mass matrix and is proportional to the up--quark mass
matrix due to the underlying $SO(10)$ symmetry  \cite{tauneutrinomass}. 
Taking the mass matrices to be nearly diagonal the physical eigenvalues are
\beq
m_{\nu_j} \sim  
\left(
{{k M^j_u} \over {\langle{\overline{\cal N}}\rangle  }    }
\right)^2 \langle{\phi}\rangle~~~,~~~
\qquad m_{N_j},m_{\phi} \sim \langle{\overline {\cal N}}\rangle~.
\label{neutrinomasseigen}
\eeq
where $k$ is a renormalisation factor due to RGE evolution, and 
$ \langle{\overline {\cal N}}\rangle \gtrapprox 10^{14}$GeV, as required 
to produce sufficiently suppressed left--handed tau neutrino mass 
\cite{tauneutrinomass}. This mass scale is dictated by the $SO(10)$
relations between the top quark and tau neutrino Dirac mass terms. 
The result is that the mass eigenvalues consist of one light left--handed 
neutrino and two heavy sterile states at the 
$\langle{\overline {\cal N}}\rangle$ scale. There is no light sterile neutrino. 
This result is borne out in detailed studies of neutrino masses in 
free fermionic models \cite{nmasses, cfnmasses}. The conclusion of these 
studies is that it is hard enough to generate sufficiently suppressed 
left--handed neutrino masses, let alone any additional light sterile
neutrinos. Furthermore, we generally expect any non--chiral state
to receive mass of the order of the GUT or string scales. Only
states that are chiral with respect to some gauge symmetry remain 
remain massless down to lower scales. This outcome is supported 
by detailed analysis in concrete models. Some mass terms of
vector--like states may arise from nonrenormalisable terms, and 
therefore suppressed compared to the leading terms. They will appear
at intermediate scale, a few orders of magnitude below the GUT or Planck scales, 
but not at a scale required for sterile neutrinos. 
The conclusion is that existence of sterile neutrinos at low scales requires 
that they are chiral with respect to an additional $U(1)$ gauge symmetry
beyond the Standard Model. The extra $U(1)$ symmetry may be broken at intermediate
scale above the TeV scale, and may be within reach of contemporary experiments. 

\section{Low scale $Z^\prime$ in free fermionic models} 

The interest in string inspired $Z^\prime$ models followed from the 
observation that string inspired effective field theory models 
give rise to $E_6$ GUT like models. Extra $U(1)$ symmetries 
in these string inspired models therefore possess an $E_6$ embedding 
and have generated multitude of papers since the mid--eighties. However, 
the construction of string derived models that admit an unbroken 
extra $U(1)$ symmetry down to low scales proves to be more of a challenge. 
The combination of $U(1)_{B-L}$ and $U(1)_{T_{3_R}}$ given by
$$U(1)_{Z^\prime} ~=~ {3\over 2} U(1)_{B-L} -2 U(1)_{T_{3_R}}$$
is embedded in $SO(10)$ and orthogonal to the weak hypercharge 
combination $U(1)_Y= {1\over 2} U(1)_{B-L} + U(1)_{T_{3_R}}$ \cite{so10zprime}.
Preserving this combination down to low scales guarantees that dimension
four operators that may induce proton decay are sufficiently suppressed 
\cite{so10zprime}. However, as discussed above the resulting seesaw
neutrino mass scale is too low to produce sufficient suppression of 
left--handed neutrino masses \cite{tauneutrinomass},
On the other hand, the symmetry breaking pattern in the string 
models $E_6\rightarrow SO(10)\times U(1)_A$ entails that $U(1)_A$ 
is anomalous and cannot be part of a low scale unbroken $U(1)_{Z^\prime}$
\cite{u1a}. String derived models with low scale 
$U(1)_{Z^\prime}\notin E_6$ were studied in \cite{none6zprime}.
However, agreement with the measured values of $\sin^2(\theta)_W(M_Z)$ and
$\alpha_s(M_Z)$ favours $Z^\prime$ models with $E_6$ embedding 
\cite{fmzpgcu}. A $Z^\prime\in E_6$ string derived model was presented in 
\cite{frzprime}. 

The construction of the string derived $Z^\prime$ model in \cite{frzprime}
utilises the spinor--vector duality \cite{svd} that was observed in the
classification of free fermionic heterotic--string $SO(10)$ models
\cite{so10class}. The classification was extended to models in which the 
$SO(10)$ symmetry is broken to 
$SO(6)\times SO(4)$ \cite{psclass};
$SU(5)\times U(1)$ \cite{fsu5class};  
$SU(3)\times SU(2)\times U(1)^2$ \cite{slmclass}; 
and 
$SU(3)\times U(1) \times SU(2)^2$ \cite{lrsclass}.
The free fermionic classification method uses a fixed set of boundary condition 
basis vectors and the enumeration of vacua is obtained by varying the GGSO
projection coefficients of the one loop partition 
function\footnote{See {\it e.g.} talk by Benjamin Percival at this conference}.
The basis typically contains thirteen or fourteen boundary condition basis
vectors. Correspondingly there are up to 91 modular invariant independent 
binary phases. Some of the free phases are fixed by physical requirements, 
such as the requirement that the vacua possess $N=1$ spacetime supersymmetry.
The space of scanned vacua therefore correspond to some $10^{15}$ 
independent $Z_2\times Z_2$ orbifold heterotic--string vacua. 
The entire physical spectrum is analysed by expressing the 
GGSO projection conditions of the massless states arising in all the states
producing sectors in the form of algebraic equations, which are coded in
a computer program. In this manner a large space of models and 
their entire physical spectrum is spanned and analysed according to 
prescribed physical criteria. 

An example of the utility of the method is provided by 
spinor--vector duality \cite{svd}. The duality is under the 
exchange of the total number of $(16+\overline{16})$ representations
of $SO(10)$ with the total number of $10$ representations. For every model
with a number $\#_1$ of $(16+\overline{16})$ and $\#_2$ of $10$ there is 
another models with a $\#_2$ of $(16+\overline{16})$ and $\#_1$ of $10$. 
The duality is easy to understand if we consider the extension of 
$SO(10)\times U(1)_A$ to $E_6$. The chiral and anti--chiral representations 
of $E_6$ decompose under $SO(10)\times U(1)$ as $27=16+10+1$ and 
$\overline{27}= \overline{16}+10+1$. In this case $\#_1=\#_2$. 
The $E_6$ enhanced symmetry 
point therefore correspond to the self--dual point under the exchange 
of the total number of $SO(10)$ spinor plus anti--spinor, with the 
total number of vectors. The compactifications with $E_6$ symmetry 
correspond to heterotic--string compactifications with $(2,2)$ worldsheet
supersymmetry, and the breaking of $E_6\rightarrow SO(10)\times U(1)_A$ 
correspond to the breaking of the $(2,2)$ worldsheet supersymmetry to
$(2,0)$ (or $(2,1)$ \cite{u1a}). The important observation is that the 
spectral flow operator of the right--moving $N=2$ worldsheet supersymmetry
is the operator that induces the map between the spinor--vector dual vacua. 
This duality was initially observed from the classification of the large
spaces of fermionic $Z_2\times Z_2$ orbifold and subsequent deeper 
understanding ensued in terms of the worldsheet properties. 

The classification method provides a fishing method to extract models 
with specified physical properties. In order to remain unbroken down 
to low scales an extra $U(1)$ symmetry has to be anomaly free.
The $Z_2\times Z_2$ orbifold allows the existence of self--dual models 
under spinor--vector duality, in which the gauge symmetry is, however, 
not enhanced to $E_6$. In these models the total number of
$(16+\overline{16})$ is equal to the total number of $10$ 
representations. In such models the chiral spectrum still forms 
complete $E_6$ representations, but the gauge symmetry is not 
enhanced to $E_6$. This is possible in the $Z_2\times Z_2$ 
orbifolds if the different components that form complete $E_6$ 
multiplets arise from different fixed points of the $Z_2\times Z_2$ 
orbifold. That is if both a spinorial and vectorial states are massless
at the same fixed point, the symmetry is necessarily enhanced to $E_6$. 
However, if they are obtained from different fixed points, the spectrum
may be self--dual and form complete $E_6$ multiplets without enhancement
of the $SO(10)\times U(1)_A$ 
symmetry to $E_6$. In this case, by virtue of the fact that 
the massless states form complete $E_6$ multiplets $U(1)_A$ is anomaly free
and may remain unbroken to low scales. In ref. \cite{frzprime} such 
a spinor--vector self dual model was extracted with subsequent breaking 
of the $SO(10)$ symmetry to $SO(6)\times SO(4)$, which preserves the 
spinor--vector self--duality. This model is a string derived model
in which an extra $U(1)$ with $E_6$ embedding may remain unbroken down to
low scales. Alternative scenarios to the self--dual string derived model of 
ref. \cite{frzprime} are models with different $E_8$ symmetry breaking patterns
\cite{e8patterns}. 

The extra $U(1)_{Z^\prime}$ symmetry requires the existence of additional matter 
states that obtain mass at the $U(1)_{Z^\prime}$ breaking scale. The 
additional matter states ensure that $U(1)_{Z^\prime}$ is anomaly free. The
chiral spectrum forms complete $E_6$ multiplets, which includes an $SO(10)$ 
singlet state that can serve as a sterile neutrino. The neutrino mass matrix
for one generation  takes the form

\begin{equation}
~~~~~~~~~~~~~~~~~~~~{\bordermatrix{
    ~~           & L_i       & S_i           &  H_i        & \overline{H}_i   & N_i   \cr
L_i~~            &  0        &       0       &  0          & \lambda n       & \lambda v    \cr
S_i~~            &  0        &       0       & \lambda v_2 & \lambda v_3      & 0     \cr 
H_i~~            &  0        &  \lambda v_2  &  0          & z^\prime          & 0     \cr
\overline{H}_i~~ &  \lambda n&  \lambda v_3  & z^\prime      & 0               & 0     \cr
N_i~~            &  \lambda v&       0       &  0          & 0               & {{{\cal N}}^2}/{M} 
\cr}},
\nonumber\\
\nonumber\\
\label{sterilematrix}
\end{equation}
A plausible numerical scenario per generation is given by 
\beqn
\lambda v   &=& 1{\rm GeV};~~~~~~~~~~~~~~~~~~
\lambda v_2 = 5\times 10^{-4}{\rm GeV}\approx m_e; 
~~~~~~~~\lambda v_3 = 5\times 10^{-4}{\rm GeV} \approx m_e;\nonumber\\
\lambda n   &=& 5\times 10^{-4}{\rm GeV} \approx m_e;~
~
~ 
z'     =  5\times 10^4{\rm GeV} = 50{\rm TeV};~
~~~~~~\overline{\cal N} = 5\times10^{14}{\rm GeV},~\nonumber
\eeqn
In this scenario there are two light eigenstates which are 
mixture of $\nu_L^i$ and $S_i$, with $\sin\theta\approx 0.98$ and $m_{1,2}~~
=\{10^{-2}$eV, $10^{-3}$eV\}; two nearly degenerate eigenstates 
of $H_i$ and ${\bar H}_i$ with 
$m_{3,4} = \{50$TeV, 50TeV, \} and a heavy eigenstate $N_i$ with 
$m_5 = 2.5\times 10^{11}$GeV.

\section{Conclusions}

Sterile neutrinos are hard to accommodate in heterotic--string derived 
models, unless their lightness is protected by an additional $U(1)$ gauge 
symmetry under which they are chiral. Further substantiation of the 
observations of the 
LSND and MiniBooNE collaborations may 
imply the existence of additional matter and interaction at a scale 
within reach of contemporary colliders, ushering a new 
era of discovery. Point quantum field theory models will continue to be used 
to parametrise the experimental data. However, 
the synthesis of gravity with the gauge interactions mandates a departure 
from the point idealisation of elementary particles. String theory models, 
of which the fermionic $Z_2\times Z_2$ orbifolds serves as a fertile 
crescent over the past 30 years, provide a self consistent framework 
to explore this synthesis.

\medskip
{\bf Acknowledgments}

I would like to thank the Weizmann institute and Tel Aviv university for
hospitality.

\end{document}